# Tunable Surface Phonon Polariton based THz Cavity

**MOHSEN JANIPOUR,** [1,*] **MATTHIAS HENSEN,** [2] **AND WALTER PFEIFFER,** [1]

[1] *Fakultät für Physik, Universität Bielefeld, Universitätsstr. 25, Bielefeld 33615, Germany*
[2] *Institut für Physikalische und Theoretische Chemie, Universität Würzburg, Am Hubland, 97074 Würzburg, Germany*
*\*mjanipour@physik.uni-bielefeld.de, pfeiffer@physik.uni-bielefeld.de*

**Abstract:** Designing and realizing switchable cavities at the infrared and THz frequencies is crucial for achieving novel integrated photonic circuits. Semiconductors like SiC support the propagation of surface phonon polaritons with the ability to tune carrier concentrations via doping. In this manuscript, we investigate the tunability of coupled surface-phonon-plasmon polariton waves inside an elliptical cavity build from a metal-semiconductor heterostructure. We theoretically study the impact of the extra *n*- and *p*-carrier doping on the resonant modes of the cavity and we show that for certain cavity modes a Purcell factor of 180 and an on/off-switching behavior can be achieved via charge injection. The cavity mode tuning range spans approximately 13 times the resonance width.

## 1. Introduction

Designing tunable optical cavities and resonators, especially for applications in the terahertz (THz) and infrared (IR) frequency regime, has attracted significant interest [1–9]. Among the several types of optical cavities in the visible regime, the surface plasmon polariton (SPP) [10] and surface-phonon-polariton (SPhP) based cavities can exhibit subwavelength size, and are tunable due to the material properties in this frequency range [4,11,12]. Especially SPP-based elliptical cavities are of particular interest since they are promising for the realization of periodical energy transfer between quantum emitters positioned in the largely separated focal spots [13,14], and for establishing entanglement between the emitters [14]. Furthermore, it has been shown that an elliptical cavity acts as an directional antenna, enabling controlled coupling of quantum emitters to the far-field under specific emission angles [9]. On the downside, plasmonic cavities studied up to now lack tunability, i.e., the cavity resonances rely on geometry only. In the THz frequency range semiconductors are appropriate candidates resonator materials because they can support low-loss SPhP waves in the IR and THz range and can be used as excellent alternatives for metals at these frequencies [16]. In fact, semiconductors such as silicon-carbide (SiC) and GaAs exhibit negative permittivity the Reststrahlen band of the THz frequency range [17,18]. In this manuscript, we demonstrate highly-tunable heterostructure elliptical cavities using a SiC substrate with different doping concentrations. Based on an optimized cavity design a relative tuning range of 13 times the resonance linewidth is achieved making the device interesting for future applications in nanophotonics. The prospect of tunability via carrier injection in suitable junctions is discussed.

The manuscript is structured as follows: First, we introduce the specific cavity design and the emerging cavity modes. Then, we discuss the optical properties of SiC cavity substrate and the impact of the extra doping on these properties in THz regime. In particular, we demonstrate the impact of charge carrier doping on the optical tunability of the cavity modes. In the last section, the obtained results of the structure are summarized and a possible future application is discussed.

## 2. Cavity design and SiC optical properties

Lattice vibrations inside semiconductors produce transverse optical (TO) and longitudinal optical (LO) resonances in a frequency band known as the Reststrahlen band, which features a negative permittivity [19]. As a result of the coupling between optical phonon resonances and the incident THz field at the semiconductor-dielectric interface and under the phase-matching conditions, the SPhP waves can be excited similar to the SPP excitation at the visible frequencies. Whereas SPPs on a gold [20–22] or silver surface exhibit lifetimes on the order of tens of femtoseconds SPhPs on semiconductor surfaces can live up to 100 ps [19] due to low-loss optical properties of the Reststrahlen band [23]. The relative longevity of SPhPs also holds when expressing life times in terms of the dimensionless $Q$-factor, i.e., the number of oscillations until the system has decayed: For SPPs on metal surfaces $Q$-factors on the order of 30 and 40 are encountered for Au and Ag [24], respectively. In the Restrahlen band of SiC SPhP $Q$-factors on the order of 300 are achieved.

The combination of these interesting optical properties of SiC and diverse SPP cavity design schemes form the basis for a novel cavity design that can find application in the research field of planar THz antennas, THz switching devices, and for efficient intracavity energy transfer. As demonstrated for plasmonic elliptical cavities they act as efficient directional antennas [15] and hence allow for efficient coupling between incident light and local intracavity nanoantennas [25–30]. Here, we extend these concepts towards THz frequencies and show that the cavity modes are tunable over a substantial frequency range. In the next section, we study an Ag-SiC heterostructure cavity configuration and its optical response determined by the SPhP dispersion function.

## 2.1. Cavity configuration

In this part, we present an elliptical cavity in which the excited polaritonic surface waves form standing wave patterns. Figure 1(a) shows the heterostructure cavity consisting of an elliptical cutout in an Ag film which is deposited on a SiC substrate. The ellipse has the dimensions of $a = 32\,\mu m$, $b = 10\,\mu m$, drilled in a thick Ag film by using, e.g., focused ion beam milling or conventional photolithography. Since the penetration depth of the electric field into the air halfspace is about $1.5\,\mu m$ the thickness of the Ag layer is chosen to be $h = 15.4\,\mu m$ so that the mode energy is confined inside the cavity.

Throughout this manuscript a commercial finite-difference time-domain (FDTD) solver is utilized (FDTD Solutions 8.19.1438, Lumerical Solutions Inc.) in order to numerically simulate the proposed cavity. In these simulations the unit cell dimensions are set to $\Delta x = \Delta y = \Delta z = 50$ nm and no mesh refinement is applied to the computational region. Due to the large computational region the simulation time is set to 23 ps with a time step of $\Delta t = 0.19$ fs which satisfies the Courant-Friedrichs-Lewy (CFL) stability condition of $\Delta t \leq (c\sqrt{\Delta x^{-2} + \Delta y^{-2} + \Delta z^{-2}})^{-1}$, in which $c$ defines the vacuum speed of light.

The cavity modes are excited by a an electric dipole which oscillates with a central frequency of $\omega_0 = 166.4 \times 10^{12}$ rad/s perpendicular to the substrate and is positioned at one focal point of the cavity ($f_1$), 200 nm above the SiC surface. The dipole source operates in the Reststrahlen band of SiC in order to excite SPhP waves at the air-SiC interface. The eigenmodes of the elliptical cavity can be described by the angular and radial Mathieu functions, in which the roots, i.e., the resonant modes, are categorized in the form of (*m*, *n*). Here, *m* and *n* denote the number of nodal lines in the field distribution along the *x*-axis and *y*-axis, respectively.

Figure 1(b) shows the normalized intensity and phase of the *z*-component of the electric field at the focal point $f_2$. We identify four resonance peaks in the Reststrahlen band. Accordingly, the (9,1)-, (10,1)-, (11,1)-, (12,1), and (13,1)-mode occurs at $\omega$ = 151.0, 159.5, 167.7, 174.7, 180.4 $\times 10^{12}$ rad/s, respectively. At lower and higher frequencies the modes are less pronounced. This is due to an increasing imaginary part of permittivity $\varepsilon$ and the epsilon-near-zero behavior of the real part of $\varepsilon$ at low and high frequencies, respectively (see Figure 2a,b). Figures 1(c) and 1(d) depict intensity cross sections of the (10,1)-mode in the *xz*- and *xy*-plane, respectively, for a carrier concentration of $n = 10^{20}\,cm^{-3}$. These cross sections show that the excitation of a resonant mode pattern leads to an increased intensity at the conjugated focal point $f_2$. Since the electric field intensity distribution is rapidly damped inside the SiC substrate, and because of its confinement at the air-SiC interface, the emerging mode can be considered as a resonant surface wave.

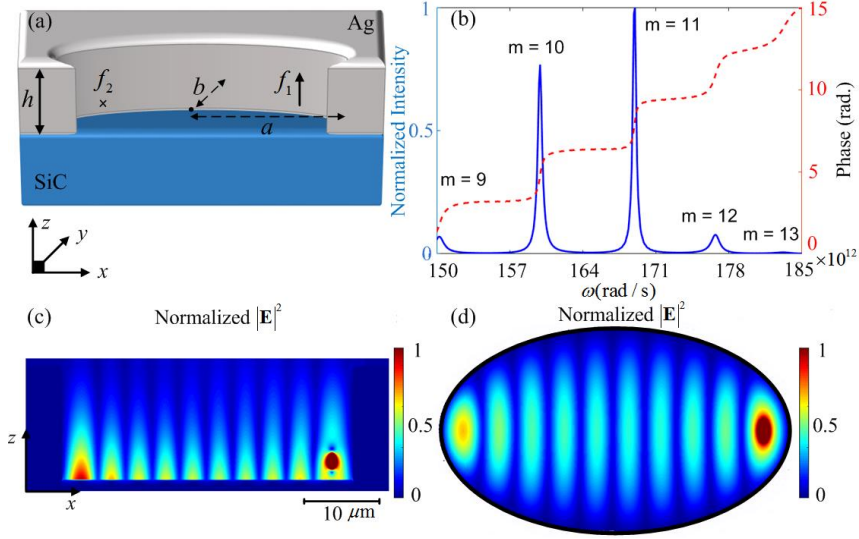

Figure 1. (a) Schematic representation of the proposed cavity structure. (b) The spectral electric field intensity (solid-curve) and phase (dashed curve) for heavily doped SiC (carrier concentration of $n = 10^{20}\,\text{cm}^{-3}$). (c), (d) Normalized electric field intensity distribution for heavily doped SiC with $n = 10^{20}\,\text{cm}^{-3}$ at $\omega = 159.5 \times 10^{12}$ rad/s in the xz- and xy-cross-sections in (c) and (d), respectively.

## 2.2. SiC optical properties

### 2.2.1. n-doped SiC

The recent progress in fabrication processes featuring sublimation growth and diffusion techniques facilitates low defect n-doping of SiC using N, P, and Ge. Carrier concentrations ranging from $n = 10^{18} - 10^{20}\,\text{cm}^{-3}$ are achievable [31]. In combination with SPhPs the ability of doping SiC has initiated research on nanophotonic devices [18,32]. The optical response of SiC can be described by a dispersive and uniaxial permittivity tensor $\varepsilon_{\perp,\parallel}(\omega) = [U]_{3\times 3}[\varepsilon_\perp\ \varepsilon_\perp\ \varepsilon_\parallel]$, in which $[U]$ is the identity matrix [33,34]. Since $\varepsilon_\perp$ and $\varepsilon_\parallel$ differ only by a small amount, we here neglect the anisotropy of the dielectric response and work with an isotropic response function. Since the dipole source oscillates perpendicular to the substrate the perpendicular permittivity $\varepsilon_\perp$ of 4H-SiC is considered and, to simplify the notation, the term $\varepsilon(\omega,n)$ is used throughout the manuscript. The optical response of SiC is affected by the contribution of phonons and the doping-influenced plasma frequency, which in turn can be explained by Lorentz, and Drude classic oscillator models, respectively. Thus, the dielectric function of SiC can be written as $\varepsilon(\omega,n) = \varepsilon_{Lorentz}(\omega) - \varepsilon_{Drude}(\omega,n)$ with [35]:

$$\varepsilon_{Lorentz}(\omega) = \varepsilon_\infty\left[1 + \frac{\omega_{LO}^2 - \omega_{TO}^2}{\omega_{TO}^2 - \omega(\omega + i\Gamma)}\right] \text{ and } \varepsilon_{Drude}(\omega,n) = \frac{\omega_p^2}{\omega[\omega + i\gamma(n)]} \text{ with } \omega_p(n) = \sqrt{\frac{n e^2}{m_n^* \varepsilon_0 \varepsilon_\infty}}, \quad (1)$$

where $\varepsilon_\infty = 9.6$ is the high-frequency permittivity, $\Gamma = 0.52 \times 10^{12}$ rad/s defines the phonon damping, and $\omega_{TO} = 150 \times 10^{12}$ rad/s and $\omega_{LO} = 180 \times 10^{12}$ rad/s are the transverse optical (TO) and longitudinal optical (LO) phonon angular frequency, respectively [19,36]. Note that since the number of the free charge carriers is approximately zero, i.e., $n = 0$, the Lorentzian contribution to the dielectric function is dominant for undoped SiC. For the n-doped cases the electron collision rate can be written as $\gamma(n) = e(m_n^* \mu(n))^{-1}$, where $e$ is the elementary charge, $m_n^* = 0.37 m_0$ is the effective mass, and $\mu(n) = 1020 \left(1 + 5.55 \cdot 10^{-18} n^{0.6}\right)^{-1}$ cm$^2$V$^{-1}$s$^{-1}$ is the electron mobility of the dopants [33,35,37,38].

According to Eq. (1), an incident THz field under phase-matching conditions can excite the surface waves which can be interpreted as the hybrids of SPhPs and SPPs (or in short SPhP/SPP) waves [18,39]. Figures 2(a) and 2(b) illustrate the real and imaginary parts of the doped SiC bulk medium for different doping concentrations, respectively. The frequency band between $\omega_{TO}$ and $\omega_{LO}$ defines the Reststrahlen band which supports SPhP/SPP waves at the SiC-vacuum interface. As can be seen in Fig. 2(a) a higher doping shifts $\omega'_{LO}$ to higher frequencies and thus leads to a broadening of the Reststrahlen band. Moreover, Fig. 2(b) shows that increasing the doping results in a larger imaginary part of $\varepsilon(\omega,n)$ due to higher collision rates.

Due to the non-zero imaginary part of $\varepsilon(\omega,n)$ these waves decay while they propagate along the interface. Thus, studying the effect of carrier concentration on their propagation length $l_{SPhP}$ is useful for designing novel devices. Figure 2(c) shows $l_{SPhP}$ as a function of carrier concentration and frequency. Note that the doping concentration mainly affects the mode tunability at frequencies slightly below $\omega'_{LO}$. For instance, consider a doping of $10^{19}$ cm$^{-3}$ and a spectral region close to the (13,1)-mode, i.e., the region marked by the black square in Fig. 2(c). Furthermore, Fig. 2(c) reveals that the SPP/SPhP waves can exhibit a propagation length between 10-100 $\mu$m in the defined window near $\omega'_{LO}$.

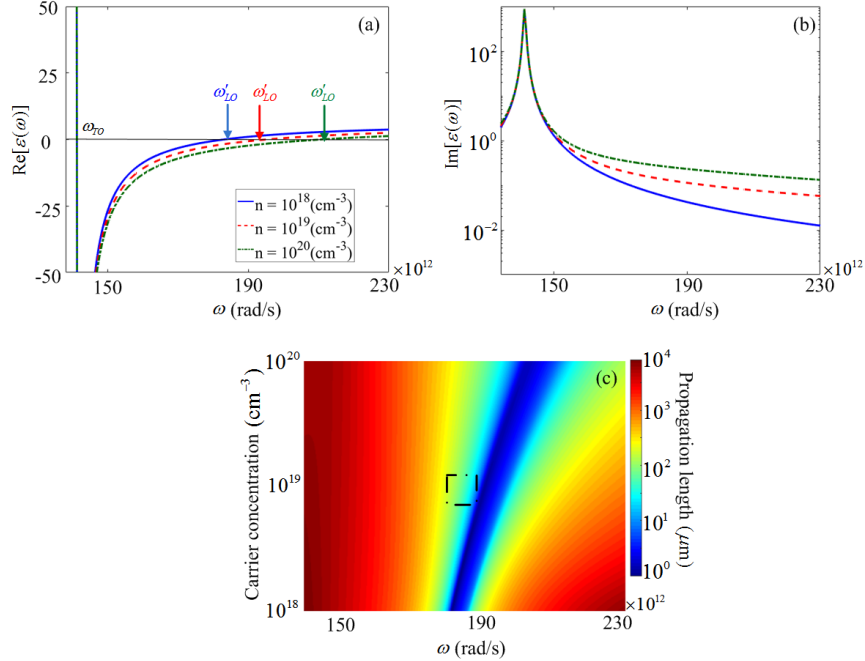

Figure 2. (a) Real, and (b) imaginary parts of the dielectric function of SiC for $n = 10^{18}$ cm$^{-3}$ (solid line), $n = 10^{19}$ cm$^{-3}$ (dashed), and $n = 10^{20}$ cm$^{-3}$ (dash-dotted), respectively. (c) The propagation length $l_{SPhP}$ $\mu$m of the SPhP/SPP modes at the air-SiC interface versus carrier concentration (cm$^{-3}$). The dashed-dotted window shows a sample part of the desired $l_{SPhP}$ spectra.

To study the tunability and the resolution of the cavity modes via the carrier concentration we have investigated the mode spectrum as a function of the doping level (see Fig. 3(a)). The (9,1)-mode at $151 \times 10^{12}$ rad/s, i.e., close to $\omega_{TO}$, is almost unaffected by the doping level, whereas cavity modes still within the Reststrahlen band, and close to $\omega'_{LO}$, behave differently: Their peaks show height variations and substantial spectral shifts. For example the (11,1)-mode exhibits a significant blue shift from 165 rad/s to $167.7 \times 10^{12}$ rad/s and an increase of the Purcell factor (see Fig. 3(b)), although the loss is increased by 100% as the carrier density increases from $n = 10^{18}$ cm$^{-3}$ to $n = 10^{20}$ cm$^{-3}$. The corresponding maximum field intensity at the focal point in the cavity is shifted in frequency and varies non-monotonously. This behavior is attributed to a change in both the real and imaginary part of the dielectric material as the doping level is increased. Note that close to the high frequency limit of the Reststrahlen band the (12,1)-mode only appears at the highest doping level since at lower doping $l_{SPhP}$ is just too short to allow the formation of a mode inside the cavity.

As a criterion to quantify the tunability of the local field strength in the cavity one can employ the Purcell factor, which compares the power loss of a dipolar source inside the cavity structure to the spontaneous decay rate of the same dipole placed in vacuum. Figure 3(b) shows the Purcell factor calculated using the built-in function of the simulation software. According to Fig. 3(b), the Purcell factor and the field intensity (Fig. 3(a)) behave very similar. Again, increasing the doping values from $n = 10^{19}$ cm$^{-3}$ to $n = 10^{20}$ cm$^{-3}$ significantly shifts the spectral features in the limit of the Reststrahlen band as it is indicated by colored bars in Fig. 3(b). It should be noted that although the shifted $\omega'_{LO}$ for each individual doping level is much higher, the tunability behavior and the resonant modes can be observed in the spectral domain shown in Figs. 3(a) and 3(b). The tunability can be quantified by $\delta = (\omega_{Shifted} - \omega_{Original}) BW_{BW}^{-1}$, where $\omega_{Shifted}$, $\omega_{Original}$, and $BW_{FWHM}$ refer to the original and shifted resonant mode frequencies, and the full-width half-maximum (FWHM) bandwidth of the mode at $\omega_{Shifted}$, respectively. According to Fig. 3(b) the mode tunability in units of the mode bandwidth reaches $\delta = 13$ for the (11,1)-mode as the doping level is increased from $n = 10^{18}$ cm$^{-3}$ to $10^{20}$ cm$^{-3}$, whereas the Purcell factor remains almost unchanged at a large value of about 180. This amount of $\delta$ together with a constant Purcell factor show that by tuning the carrier level, the (11,1)-mode can be purely switched on/off at its individual frequency.

For some applications it is important to achieve a linearized frequency shift among neighboring modes. Therefore, as a measure of the linearity of the mode shifting, one can define the spectral separation between the resonant modes for each doping as $\Delta\omega = \omega_{m+1} - \omega_m$, where m identifies the mode according to (m, n). The doping-dependent spectral separation $\Delta\omega$ is shown in the inset of Fig. 3(b). Indeed, the right choice of carrier concentration can improve the linearity of $\Delta\omega$, such that for the heavily doped case of $n = 10^{20}$ cm$^{-3}$ we acieve an almost constant frequency spacing of around $\Delta\omega = 8.8 \times 10^{12}$ rad/s.

Practically, the tuning of the SPhP-SPP interaction in SiC via carrier concentration can be realized by photocarrier injection. This method is based on the coupling of photoinjected carriers and SPhPs, resulting in a shift of the SPhP resonance frequency using UV excitation pulses [40]. Thus, active tuning would be possible either optically by injecting the photocarriers or electrically by applying a bias voltage to the heterostructure, which results in a modification of the number of free charge carriers.

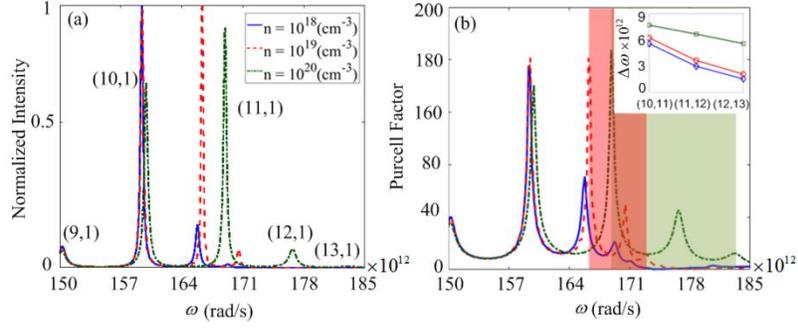

Figure 3. (a) The normalized electric field intensity at $f_2$ point and the (b) the Purcell factor of the excited cavity through the electric dipole at point $f_1$ for $n = 10^{18}\,\text{cm}^{-3}$ (solid line), $n = 10^{19}\,\text{cm}^{-3}$ (dashed), and $n = 10^{20}\,\text{cm}^{-3}$ (dash-dotted), respectively. The inset shows the spectral spacing between the different modes, i.e., $\Delta\omega = \omega_{m+1} - \omega_m$ where $m = 10$, 11, 12, and 13 for $n = 10^{18}\,\text{cm}^{-3}$ (blue-diamond), $n = 10^{19}\,\text{cm}^{-3}$ (red-circle), and $n = 10^{20}\,\text{cm}^{-3}$ (green-square), respectively.

### 2.2.2. p-doped SiC

Doping free $p$-carriers in SiC can be realized by ion implantation of Al and B in which the implanted layers can provide very high acceptor concentrations in a broad temperature range [37,38]. Similar to the $n$-doped SiC the optical response of the $p$-doped SiC can be explained by the previously mentioned dispersive, and uniaxial, permittivity tensor (see Section 2.2.1). Although the Lorentzian oscillator model remains unchanged for the $p$-doped SiC material the Drude part of Eq. (1) needs to be modified. The hole collision rate can be written as $\gamma(p) = e\left(m_p^* \mu(p)\right)^{-1}$, where $e$ is the elementary charge, $m_p^* = 0.66\,m_0$ the effective mass, and $\mu(p) = 118\left(1 + 4.54 \cdot 10^{-19} p^{0.7}\right)^{-1}\,\text{cm}^2\,\text{V}^{-1}\,\text{s}^{-1}$ is the mobility of the holes [41]. Figures 4(a) and 4(b) illustrate the real and imaginary parts of $\varepsilon(\omega, n)$ for $p$-doped SiC with different $p$-type carrier concentrations, respectively. In contrast to $n$-type doping, the hole concentration does not substantially influence the real part of the dielectric function since the Drude term in $\varepsilon(\omega, n)$ plays only a minor role due to the low mobility of holes. Nonetheless, an increasing $p$-type carrier concentration introduces a slight blue-shift of $\omega_{LO}$, as shown in the inset of Fig. 4(a). However, the dominating effect is an increasing imaginary part of $\varepsilon(\omega, n)$, especially for the heavily doped cases.

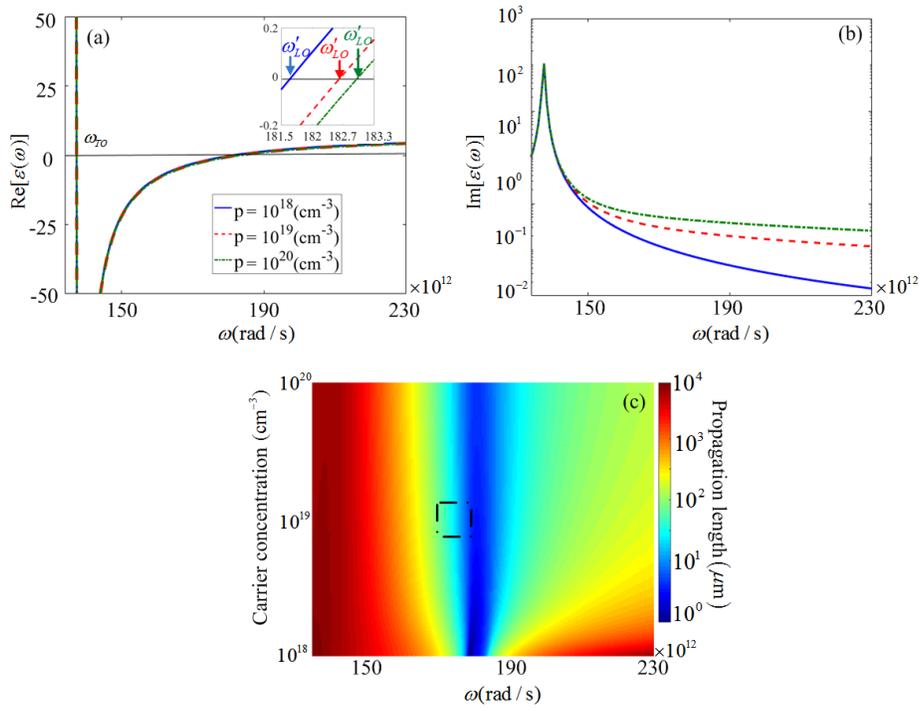

Figure 4. (a) Real, and (b) imaginary parts of the dielectric function of SiC for $p = 10^{18}\,\text{cm}^{-3}$ (solid line), $p = 10^{19}\,\text{cm}^{-3}$ (dashed), and $p = 10^{20}\,\text{cm}^{-3}$ (dash-dotted), respectively. (c) The propagation length $l_{\text{SPhP}}$ of the SPhP/SPP waves at the air-SiC interface versus $p$-carrier concentration.

Strictly speaking, the collision rate of holes in the doped SiC substrate is approximately four times higher than for electrons, i.e., $\gamma(p) \approx 4\gamma(n)$, and it hence significantly increases the amount of loss. This behavior has a smaller effect on the real part of the dielectric function $\varepsilon(\omega, p)$ compared to $n$-type doping and hence a very small shift of $\omega_{LO}$ is observed. Figure 4(c) depicts the propagation length of the SPhP/SPP waves at the air-SiC boundary as a function of the $p$-doped carrier concentration. It can be seen that in contrast to the $n$-doped case, in which the frequencies corresponding to the desired $l_{\text{SPhP}}$ are tilting to the higher frequencies by increasing the carriers, the frequencies favorable to the maximum propagation length stay around or slightly smaller than $\omega_{LO}$ (see the dashed-dotted area).

Figures 5(a), and 5(b) show the normalized electric field intensity and the Purcell factor of the cavity for different doping levels, respectively. The (9,1)-, (10,1)-, and (11,1)-modes can be clearly identified in the field intensity representation, whereas the (12,1)-mode is only visible in the Purcell factor spectrum. Interestingly, the (11,1)- and (12,1)-mode for $p = 10^{19}\,\text{cm}^{-3}$ exhibit higher field intensities and Purcell factors in comparison with the scenarios of lower and higher doping concentrations. This is attributed to the fact that for $p = 10^{19}\,\text{cm}^{-3}$ the propagation length associated frequency around $\omega_{LO}$ is more tilted to the higher frequencies. The lower propagation length along with the negligible displacement of the $\omega_{LO}$ significantly reduces the electric field intensity for the (13,1)-mode.

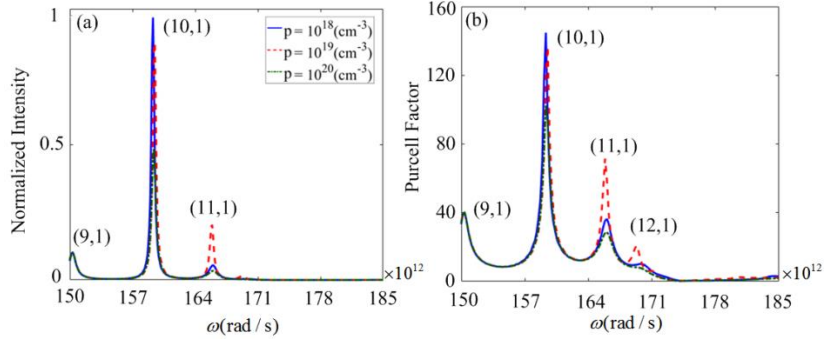

Figure 5. (a) The electric field intensity and (b) The Purcell factor of the cavity for $p = 10^{18}\,\text{cm}^{-3}$ (solid line), $p = 10^{19}\,\text{cm}^{-3}$ (dashed), and $p = 10^{20}\,\text{cm}^{-3}$ (dash-dotted), respectively.

## Conclusion

In conclusion, we have designed and demonstrated a tunable elliptical heterostructure cavity in the THz frequency range based on surface polariton waves. We have studied the effect of *n*- and *p*-carrier doping on the cavity properties. We have shown that by doping with *n*-carriers certain resonant modes can be spectrally tuned with an almost constant spectral separation between adjacent modes. Shifting of cavity modes over up to 13 times the mode bandwidth is achieved via *n*-type doping of the SiC substrate. In addition, our results show that the Purcell factor for these cavity modes reaches values of almost 200 for emitters positioned close to the focal points of the cavity. Interestingly, the Purcell factor increases for the (11,1)-, (12,1)-, and (13,1)-modes with increasing *n*-type carrier concentration, although the damping increases in the material. The large tunability of the here demonstrated cavities could be realized via laser-pulse—induced photocarrier injection. We anticipate applications of the presented structure for example in planar phased array antennas in in the THz range.

## Acknowledgments

This material is supported by Georg Forster Research Fellowship program by the Alexander von Humboldt Foundation.